# THE ANGULAR POWER SPECTRUM OF BATSE 3B GAMMA-RAY BURSTS[1]

Max Tegmark[1,2], Dieter H. Hartmann[3], Michael S. Briggs[4] & Charles A. Meegan[5]

[1] Max-Planck-Institut für Physik, Föhringer Ring 6, D-80805 München;
max@mppmu.mpg.de

[2] Max-Planck-Institut für Astrophysik, Karl-Schwarzschild-Str. 1, D-85740 Garching

[3] Dept. of Physics and Astronomy, Clemson University, Clemson, SC 29634;
hartmann@grb.phys.clemson.edu

[4] Dept. of Physics, University of Alabama, Huntsville, AL 35899

[5] NASA/Marshall Space Flight Center, Huntsville, AL 35812

## Abstract

We compute the angular power spectrum $C_\ell$ from the BATSE 3B catalog of 1122 gamma-ray bursts, and find no evidence for clustering on any scale. These constraints bridge the entire range from small scales (which probe source clustering and burst repetition) to the largest scales (which constrain possible anisotropies from the Galactic halo or from nearby cosmological large scale structures). We develop an analysis technique that takes the angular position errors into account, which enables us to place tight upper limits on the clustering down to scales $\ell \sim 60$, corresponding to a few degrees on the sky.
The minimum-variance burst weighting that we employ is graphically visualized as an all-sky map where each burst is smeared out by an amount corresponding to its position uncertainty. We also present separate band-pass filtered sky maps for the quadrupole term and for the multipole-ranges $\ell = 3 - 10$ and $\ell = 11 - 30$, so that the fluctuations on different angular scales can be separately inspected for visual features such as localized "hot spots" or structures aligned with the Galactic plane. These filtered maps reveal no apparent deviations from isotropy.

---

# 1 INTRODUCTION

The BATSE experiment has now observed more than 1100 gamma-ray bursts. The observed angular distribution is isotropic, while the brightness distribution of bursts shows a reduced number of faint events. These observations favor a cosmological burst origin. The "great debate" on the distance scale of cosmic gamma-ray bursts (GRBs) (Fishman 1995; Lamb 1995; Paczyński 1995) considered two alternatives; cosmological bursts or events that occur in an extended Galactic halo (EGH). The old paradigm of nearby Galactic neutron stars with a Population I distribution perished due to the combined observations of an isotropic angular distribution of GRBs along with reduced source counts at the faint end of the apparent flux distribution (Meegan *et al.* 1992; Briggs, *et al.* 1995). The absence of even a weak "Milky Way" band in the GRB distribution has indeed made it hard to retain the hypothesis that local neutron stars provide the underlying source population. Some recent reviews of these and related issues are given by Briggs (1995), Fishman & Meegan (1995), and Hartmann (1995).

Although no dominant anisotropies on the sky were found in the apparent sky distribution of gamma-ray bursts, even small effects might contain valuable information about the underlying sources. The detection of a small excess of events in special directions, such as nearby stars or the Andromeda galaxy, could be a unique signature of stellar or galactic halo models, respectively. For example, a small asymmetry with respect to the galactic plane might suggest a local disk origin (Hartmann, Greiner, & Briggs 1995).

Clustering of bursts beyond that expected from random alignments might be evidence of actual clustering of the sources or of repeated emission from some sources. Observation of repetition would seriously call into question the viability of those cosmological burst models that invoke unique events, such as mergers of neutron star binaries. On the other hand, a detection of the small anisotropy induced by the Earth motion relative to the CMB (Maoz 1994; Scharf, Jahoda, & Boldt 1995; but see Brainerd 1995) would constitute a convincing proof of the cosmological origin hypothesis. These various anisotropies manifest themselves on different angular scales and with different magnitudes. Galactic features would be expected to cause large-scale distortions, while burst repetition would show its effects on the scale that is typical for BATSE source localizations (in excess of 1.6° for the 3B catalog). In addition, the instrument does not sample the sky uniformly so that we expect some distortions due to the non-uniform exposure map of BATSE.

How should we analyze the angular distribution of GRBs? Since the basic null-hypothesis of isotropy states that burst directions are distributed randomly on the sky (which is the impression derived from visual inspection of GRB catalogs), then we seek tests that can efficiently find small deviations. We first search for excess of sources towards some direction or a concentration towards some plane in the sky, i.e., we seek a dipole- or quadrupole moment. It is perhaps preferable to search for such large scale anisotropies in an unbiased way by not making reference to any particular coordinate frame (Hartmann & Epstein 1990; Briggs 1993). On the other hand, such coordinate-free methods are not necessarily the most efficient ones. If a particular anisotropy is expected, then the tests should take this information into account to optimize the search efficiency. Paczynski (1990) introduced studies of the $\cos\theta$ and $\sin^2 b$ statistics, where $b$ is the galactic latitude of the GRB and $\theta$ the angle between the GRB direction and the vector pointing to the Galactic center. It is now common practice to apply both the coordinate-free and the galactic methods to the GRB distribution (Briggs 1993; Briggs et al. 1995). These dipole- and quadrupole measures were sufficient to characterize the large-scale angular properties of GRBs when sample sizes were a few hundred bursts or less. However, the BATSE experiment has now observed so many bursts that an extension of these moment methods to higher orders is now useful. In this work we use spherical harmonic analysis (SHA) to represent and interpret the angular distribution of GRBs.

It can be shown (Horack et al. 1993; Briggs, et al. 1995) that the statistical estimates of low order multipoles are not very sensitive to the angular smearing induced by statistical and systematic localization uncertainties. This is not the case for higher order multipoles, which probe the angular density field on smaller scales. We shall address this question very carefully in this work. Small angular scales may reveal important information about the nature of the GRB sources, and localization accuracy is crucial. If associated with galaxies, we expect clustering on very small scales (Hartmann & Blumenthal 1989; Lamb & Quashnock 1993). If bursts repeat, we expect clustering at $\theta=0$ (Quashnock & Lamb 1993a). Both effects are diluted by localization uncertainties (the point spread function) and apparent power is transferred from small (or zero) angular scales to a scale given by the detector response. A traditional tool for the analysis of source clustering is the angular two-point correlation function, which was first applied to GRBs by Hartmann & Blumenthal (1989). The severe reduction in correlation strength by positional smearing was demonstrated by Hartmann, Linder, &

Blumenthal (1991). The two-point correlation function is closely related to the power spectrum (e.g., Peebles 1980) (in the ideal world with no measurement errors or shot noise, one would be found to be the spherical Fourier transform of the other). However, the correlation function and the power spectrum complement each other well, since they are affected by noise in quite different ways. This makes it worthwhile to estimate both from the data, just as has become the practice with galaxy surveys. Another method relevant to the study of clustering properties is the nearest neighbor method (e.g., Scott & Tout 1989). This method, first applied to GRBs by Quashnock & Lamb (1993a), only probes angles near the scale defined by the mean angular pair separation, which decreases with increasing sample size. We do not consider NN methods in this work.

The remainder of this paper is organized as follows. In Section 2, we generalize the standard techniques of power spectrum estimation to properly take into account the location errors and the sky exposure of the BATSE catalog. In Section 3 we apply this to the 3B data set, and in Section 4 we discuss the results.

## 2 METHOD

In this section, we derive the power spectrum estimation technique that is employed in our analysis. The first subsection reviews the statistics of point processes on a sphere. This is standard material, and has been frequently discussed in the literature in connection with the problem of estimating the angular power spectrum of point sources such as galaxies or quasars (Peebles 1973, Hauser & Peebles 1973, Peebles 1980) — see Tegmark (1995) for a recent review. The extra twist, which makes the analysis of the BATSE data more challenging, is the presence of position errors. Since some bursts are more accurately localized than others, the question of how to best weight the data is somewhat subtle — this is the topic of the second subsection. After that, we present the explicit expressions for computing the power spectrum estimates from a data set, including a simple beam function model.

### 2.1 Point processes on a sphere

We model the gamma ray burst distribution as a 2D stochastic point process $n(\hat{\mathbf{r}}) = \sum_i \delta(\hat{\mathbf{r}}, \hat{\mathbf{r}}_i)$ which is a Poisson process with intensity (average point density per steradian) $\lambda(\hat{\mathbf{r}})$. Here $\delta$ denotes the Dirac delta function on the surface of the unit sphere, and the unit vectors $\hat{\mathbf{r}}_i$ correspond to the positions of the various bursts. If we had detected a nearly infinite number of bursts, then the function $\lambda(\hat{\mathbf{r}})$ would be known with great accuracy, and the only source of errors when computing its power spectrum would be cosmic variance. Since in practice we have only a finite number of bursts (in our case 1122), our estimates of $\lambda$ itself will be inexact, leading to the additional complication known as shot noise.

A Poisson process satisfies (see *e.g.* Appendix A of Feldman *et al.* 1994)

$$\langle n(\hat{\mathbf{r}}) \rangle = \lambda(\hat{\mathbf{r}}), \tag{1}$$

$$\langle n(\hat{\mathbf{r}}) n(\hat{\mathbf{r}}') \rangle = \lambda(\hat{\mathbf{r}})\lambda(\hat{\mathbf{r}}') + \delta(\hat{\mathbf{r}}, \hat{\mathbf{r}}')\lambda(\hat{\mathbf{r}}). \tag{2}$$

Here $\lambda$ is itself a random field, $\lambda(\hat{\mathbf{r}}) = \bar{n}(\hat{\mathbf{r}})[1 + \Delta(\hat{\mathbf{r}})]$, where the underlying density fluctuations $\Delta$ are modeled as a Gaussian random field. The function $\bar{n}$, which we will refer to as the *exposure function*, is thus the number of bursts per steradian expected a priori, not the number density actually observed. In other words, $\bar{n}(\hat{\mathbf{r}})$ is proportional to the exposure time in the sky direction $\hat{\mathbf{r}}$. As customary, we assume that $\langle \Delta(\hat{\mathbf{r}}) \rangle = 0$ and that the statistical properties of the field $\Delta$ are isotropic, which means that if we expand it in spherical harmonics[2] as

$$\Delta(\hat{\mathbf{r}}) = \sum_{\ell=0}^{\infty} \sum_{m=-\ell}^{\ell} a_{\ell m} Y_{\ell m}(\hat{\mathbf{r}}), \tag{4}$$

then

$$\langle a_{\ell m} a_{\ell' m'} \rangle = \delta_{\ell \ell'} \delta_{m m'} C_\ell, \tag{5}$$

where the coefficients $C_\ell$ are known as the *angular power spectrum*. There are thus two separate random steps involved in generating $n$: first the generation of the smooth field $\Delta$, then the Poissonian distribution of points. For instance, $\langle n(\hat{\mathbf{r}}) \rangle = \langle \lambda(\hat{\mathbf{r}}) \rangle = \bar{n}(\hat{\mathbf{r}})$.

---

[2]Since all our fields are real-valued, we will find it convenient to use the *real-valued* versions of the spherical harmonics throughout. These are identical to the conventional spherical harmonics $Y_{\ell m}$ as defined in, for instance, Abramowitz & Stegun (1965), except that the complex exponentials $e^{im\varphi}$ are replaced by $\sqrt{2}\sin(m\varphi)$, 1 and $\sqrt{2}\cos(m\varphi)$ for $m < 0$, $m = 0$ and $M > 0$, respectively. With this convention, the standard identities involving spherical harmonics remain unchanged except that no complex conjugation is needed. For instance, the orthogonality relation becomes simply

$$\int Y_{\ell m}(\hat{\mathbf{r}}) Y_{\ell' m'}(\hat{\mathbf{r}}) d\Omega = \delta_{\ell \ell'} \delta_{m m'}. \tag{3}$$

Given the field $n(\hat{\mathbf{r}})$, we wish to estimate the coefficients $a_{\ell m}$. We denote our estimates $\tilde{a}_{\ell m}$, and for reasons that will soon become clear, we define them as

$$\tilde{a}_{\ell m} \equiv \int Y_{\ell m}(\hat{\mathbf{r}}) \frac{n(\hat{\mathbf{r}})}{\bar{n}(\hat{\mathbf{r}})} d\Omega - \delta_{\ell 0} \delta_{m 0} \sqrt{4\pi}. \tag{6}$$

We now compute the statistical properties of these estimates. By substituting equation (1) into equation (6), we obtain

$$\langle \tilde{a}_{\ell m} \rangle = \int Y_{\ell m}(\hat{\mathbf{r}}) d\Omega - \delta_{\ell 0} \delta_{m 0} \sqrt{4\pi} = 0, \tag{7}$$

*i.e.*, the expectation values vanish. Since the expectation values of the true coefficients $a_{\ell m}$ vanish as well, this means that our estimates are unbiased. Notice that we chose to include the second term in equation (6) simply to cancel the bias arising from the monopole term $\ell = m = 0$. Using the expressions above, we find that the correlation between two multipole estimates is

$$\langle \tilde{a}_{\ell m} \tilde{a}_{\ell' m'} \rangle = \int \int Y_{\ell m}(\hat{\mathbf{r}}) Y_{\ell' m'}(\hat{\mathbf{r}}') \left[ \langle \Delta(\hat{\mathbf{r}}) \Delta(\hat{\mathbf{r}}') \rangle + \frac{1}{\bar{n}(\hat{\mathbf{r}})} \delta(\hat{\mathbf{r}}, \hat{\mathbf{r}}') \right] d\Omega d\Omega'. \tag{8}$$

Substituting equation (4) into this, and using the spherical harmonic orthogonality relation (3) and equation (5), this reduces to

$$\langle \tilde{a}_{\ell m} \tilde{a}_{\ell' m'} \rangle = \delta_{\ell \ell'} \delta_{m m'} C_\ell + \int \frac{Y_{\ell m}(\hat{\mathbf{r}}) Y_{\ell' m'}(\hat{\mathbf{r}})}{\bar{n}(\hat{\mathbf{r}})} d\Omega. \tag{9}$$

If $\bar{n}$ is merely a constant, *i.e.*, if the exposure time is the same for all parts of the sky, then the orthogonality relation will reduce the second term to simply $\delta_{\ell \ell'} \delta_{m m'} / \bar{n}$, and the various estimates $\tilde{a}_{\ell m}$ will all be uncorrelated. Since the true exposure function $\bar{n}$ for the BATSE 3B data set varies somewhat across the sky, a slight correlation will result.

We are of course also interested in estimating the angular power spectrum $C_\ell$. Defining the quantities

$$\tilde{C}_{\ell m} \equiv \tilde{a}_{\ell m}^2 - b_{\ell m}, \tag{10}$$

we thus find that they are unbiased power estimates (in the sense that $\langle \tilde{C}_{\ell m} \rangle = C_\ell$) if we choose our *bias correction* to be

$$b_{\ell m} \equiv \int \frac{Y_{\ell m}^2(\hat{\mathbf{r}})}{\bar{n}(\hat{\mathbf{r}})} d\Omega. \tag{11}$$

If $\bar{n}$ is constant, then the bias correction becomes simply $b_{\ell m} = 1/\bar{n}$, independent of $\ell$ and $m$.

It should now be clear why we divided by $\bar{n}$ in equation (6). If we had not divided by the exposure function, then our power estimator $\tilde{C}_{\ell m}$ would not have measured only what we wanted it to, *i.e.*, $C_\ell$. Rather, $\langle \tilde{C}_{\ell m} \rangle$ would also have received contributions from other multipoles $C_{\ell'}$, with $\ell \neq \ell'$. The quantities $\tilde{C}_{\ell m}$ are thus good estimates of $C_\ell$ for each $m$-value separately. To reduce error bars, we estimate the power by averaging the $\tilde{C}_{\ell m}$:

$$\tilde{C}_\ell \equiv \frac{1}{2\ell+1} \sum_{m=-\ell}^{\ell} \tilde{C}_{\ell m}. \tag{12}$$

Defining $b$ to be the average of the bias corrections $b_{\ell m}$, we find that $b$ is in fact independent of $\ell$: by substituting the spherical harmonic addition theorem (16) into equation (11) and using the fact that $P_\ell(1) = 1$, we obtain

$$b \equiv \frac{1}{2\ell+1} \sum_{m=-\ell}^{\ell} b_{\ell m} = \frac{1}{4\pi} \int \frac{d\Omega}{\bar{n}(\hat{\mathbf{r}})}, \tag{13}$$

*i.e.*, $b$ is just the spherical average of $1/\bar{n}$. This means that the coefficients $b_{\ell m}$, which would be slightly cumbersome to compute numerically, need never be computed at all, since the power estimate $\tilde{C}_\ell$ is simply the average of the $\tilde{a}_{\ell m}$-coefficients minus $b$.

## 2.2 The effect of position errors

The discussion in the previous section applies to any population of point sources on the celestial sphere, not merely gamma-ray bursts. However, analyzing the BATSE catalog involves an extra complication that is absent in, for instance, galaxy and quasar catalogs: position errors.

Let us first study the simple case where the position errors are the same for all bursts in the catalog. If the true direction to a burst is $\hat{\mathbf{r}}$, then we model the apparent direction $\hat{\mathbf{r}}'$ as a random variable whose probability distribution depends only on the angle between $\hat{\mathbf{r}}$ and $\hat{\mathbf{r}}'$. We can thus write the probability distribution as $B(\hat{\mathbf{r}} \cdot \hat{\mathbf{r}}')$ for some function $B$ that we will refer to as the *beam function*.

Above, we characterized the distribution of the true burst positions as a Poisson process with intensity $\lambda(\hat{\mathbf{r}})$, where $\lambda$ was in turn a Gaussian random field. From now on, we will let the density $n(\hat{\mathbf{r}}) = \sum_i \delta(\hat{\mathbf{r}}, \hat{\mathbf{r}}_i)$ refer not to the true burst positions but to the apparent positions. It is easy to show that this $n$ will also be a Poisson process, but with a different intensity function $\lambda$. Specifically, the apparent intensity is the true one convolved with the beam function, *i.e.*,

$$\lambda_{app}(\widehat{\mathbf{r}}) = (B \star \lambda_{true})(\widehat{\mathbf{r}}) \equiv \int B(\widehat{\mathbf{r}} \cdot \widehat{\mathbf{r}}')\lambda_{true}(\widehat{\mathbf{r}}')d\Omega'. \tag{14}$$

Thus the effect of the position errors is to smooth out sharp features in the expected burst density, which as we will see limits our ability to measure fluctuations on scales below the beam width. Let us expand the beam function in Legendre polynomials as

$$B(\widehat{\mathbf{r}} \cdot \widehat{\mathbf{r}}') = \sum_{\ell=0}^{\infty} \left(\frac{2\ell+1}{4\pi}\right) B_\ell P_\ell(\widehat{\mathbf{r}} \cdot \widehat{\mathbf{r}}'). \tag{15}$$

By using the spherical harmonic addition theorem,

$$\sum_{m=-\ell}^{\ell} Y_{\ell m}(\widehat{\mathbf{r}}) Y_{\ell m}(\widehat{\mathbf{r}}') = \left(\frac{2\ell+1}{4\pi}\right) P_\ell(\widehat{\mathbf{r}} \cdot \widehat{\mathbf{r}}'), \tag{16}$$

together with the orthogonality relation (3), we can thus write the beam function as

$$B(\widehat{\mathbf{r}} \cdot \widehat{\mathbf{r}}') = \sum_{\ell=0}^{\infty} \sum_{m=-\ell}^{\ell} B_\ell Y_{\ell m}(\widehat{\mathbf{r}}) Y_{\ell m}(\widehat{\mathbf{r}}'). \tag{17}$$

Applying the beam convolution to equation (4) and using the orthogonality relation, we thus obtain the spherical version of the convolution theorem:

$$(B \star \Delta)(\widehat{\mathbf{r}}) = \sum_{\ell=0}^{\infty} \sum_{m=-\ell}^{\ell} a_{\ell m} B_\ell Y_{\ell m}(\widehat{\mathbf{r}}). \tag{18}$$

In other words, convolution with $B$ simply corresponds to multiplying the multipole coefficient $a_{\ell m}$ by $B_\ell$.

Repeating the analysis of the previous section including position errors (replacing $\lambda$ by $B \star \lambda$), the case where $\bar{n}$ is constant[3] thus yields the simple result

$$\langle \tilde{C}_{\ell m} \rangle = B_\ell^2 C_\ell. \tag{19}$$

---

[3]When $\bar{n}$ is not constant, the $B \star (\bar{n}\Delta)$-term in addition gives rise to a weak mode coupling between the different multipoles. As discussed below, the $\bar{n}$ of the BATSE 3B data set is basically constant except for small dipole and quadrupole corrections. This means that $\tilde{a}_{\ell m}$ will pick up small contributions from $a_{\ell' m}$, where $|\ell' - \ell| \leq 2$, which is

In practice, some sources are more accurately localized than others, and we clearly want to make use of this fact to make the most of the data. Suppose that the total population, with number density $\bar{n}$, consists of a number of sub-populations with number densities $\bar{n}_i$ (so that $\sum \bar{n}_i = \bar{n}$), and that all bursts in the $i^{th}$ sub-population are equally accurately localized, as specified by a beam function $B_i$. Estimating the power spectrum can now be split into two steps:

1. Estimate $a_{\ell m}$ separately from each population, as above, and call the results $\tilde{a}_{\ell m}^i$
2. Combine these estimates into one by some weighted averaging,

$$\tilde{a}_{\ell m} = \sum W_{\ell i} \tilde{a}_{\ell m}^i. \qquad (20)$$

We obviously want the weights $W_{\ell i}$ to be larger for those populations $i$ that are better localized. Let us now determine which weighting scheme is optimal. The generalization of equation (19) to multiple populations is readily found to be

$$\langle \tilde{C}_{\ell m} \rangle = \langle \tilde{a}_{\ell m}^2 \rangle - b_\ell = \left( \sum_i B_\ell^i W_{\ell i} \right)^2 C_\ell, \qquad (21)$$

where the bias correction is

$$b_\ell \equiv \sum_i \frac{W_{\ell i}^2}{\bar{n}_i}. \qquad (22)$$

How should we choose our weights $W_{\ell i}$? First of all, to make $\tilde{C}_{\ell m}$ an unbiased estimate of $C_\ell$, we clearly want to normalize the weights so that the expression in parenthesis in equation (21) equals unity, *i.e.*, so that $\sum_i B_{\ell i} W_{\ell i} = 1$. Secondly, we want the error bars on our estimate to be as small as possible, *i.e.*, we want to minimize the variance of $\tilde{C}_\ell$. In the approximation that $\tilde{a}_{\ell m}$ is Gaussian[4], we have simply $V(\tilde{C}_\ell) = 2\langle \tilde{a}_{\ell m}^2 \rangle^2$, so that we minimize the

---

completely irrelevant for this analysis. The reason is that it is merely a second order effect: we are investigating whether there is any signal at all apart from the shot noise, and this coupling effect would only alter the *relative* level of the signal by a few percent. Thus the only instance where the anisotropy of $\bar{n}$ must be taken into account is when computing the noise bias with equation (11), since an error of a few percent in the (much larger) shot noise contribution could be of the same order as the weak signal we are trying to detect.

[4]The Gaussian approximation is good when the number of bursts is large, by the central limit theorem. It should be emphasized that even under circumstances where this approximation is poor, our $\tilde{C}_\ell$ will be a good estimate of the power spectrum — it will simply have slightly larger error bars than it would with optimal weighting.

variance by minimizing the expectation value $\langle \tilde{a}_{\ell m}^2 \rangle = C_\ell + b$. Since $C_\ell$ is independent of our weights, we thus wish to choose $W_{\ell i}$ so as to minimize the bias correction $b$, given the above-mentioned normalization constraint $\sum_i \bar{n}_i B_{\ell i} W_{\ell i} = 1$. This constrained optimization problem is readily solved by the method of Lagrange multipliers, and the solution is

$$W_{\ell i} = b_\ell \bar{n}_i B_{\ell i}, \tag{23}$$

where the minimal bias correction is

$$b_\ell = \left[\sum_i \bar{n}_i B_{\ell i}\right]^{-1}. \tag{24}$$

In summary, we have thus found our best multipole estimate to be

$$\tilde{a}_{\ell m} \equiv \frac{N b_\ell}{4\pi} \sum_i B_{\ell i} \int Y_{\ell m}(\widehat{\mathbf{r}}) \frac{n_i(\widehat{\mathbf{r}})}{\bar{n}(\widehat{\mathbf{r}})} d\Omega. \tag{25}$$

## 2.3 Power spectrum estimation in practice

For any given data set, the density field $n_i$ is just a sum of delta functions, one for each burst, so equation equation (25) reduces to

$$\tilde{a}_{\ell m} = \frac{N b_\ell}{4\pi} \sum_i B_{\ell i} \sum_{j=1}^{N_i} \frac{Y_{\ell m}(\widehat{\mathbf{r}}_j)}{\bar{n}(\widehat{\mathbf{r}}_j)}, \tag{26}$$

where $N_i$ denotes the number of bursts in the $i^{th}$ sub-population. We can simplify this expression further by a mere change of notation. We let the index $k$ refer to sums over the entire burst sample ($k = 1, ..., N$), and from here on, we simply let $B_{\ell k}$ denote the beam factor corresponding to the sub-population that the $k^{th}$ burst belongs to. Then

$$b_\ell^{-1} = \sum_i \bar{n}_i B_{\ell i}^2 = \frac{1}{4\pi} \sum_i N_i B_{\ell i}^2 = \frac{1}{4\pi} \sum_k B_{\ell k}^2 = \frac{N_\ell^{\text{eff}}}{4\pi}, \tag{27}$$

where we have defined the *effective number of bursts* at a given multipole as $N_\ell^{\text{eff}} = \sum_{k=1}^N B_{\ell k}^2$. With this same convention, replacing the double sum over sub-populations and their members by a single sum over all bursts, equation (25) simplifies to

$$\tilde{a}_{\ell m} = \frac{N}{N_\ell^{\text{eff}}} \sum_{k=1}^N B_{\ell k} \frac{Y_{\ell m}(\widehat{\mathbf{r}}_k)}{\bar{n}(\widehat{\mathbf{r}}_k)}. \tag{28}$$

Thus we have eliminated the need to keep track of sub-populations altogether[5], and expressed out multipole estimates directly in terms of the observed quantities.

Repeating the analysis for an arbitrary exposure function $\bar{n}$, equation (27) becomes generalized to

$$b_\ell = \frac{N}{N_\ell^{\rm eff}} \frac{1}{4\pi} \int \frac{d\Omega}{\bar{n}(\hat{\bf r})}. \tag{29}$$

We estimate the $C_\ell$ by averaging over $m$-values as before, $i.e.$,

$$\tilde{C}_\ell \equiv \left[ \frac{1}{2\ell+1} \sum_{m=-\ell}^{\ell} \tilde{a}_{\ell m}^2 \right] - b_\ell, \tag{30}$$

In the above-mentioned Gaussian approximation, the $\tilde{C}_{\ell m}$ of equation (10) are almost independent with variance $V[\tilde{C}_{\ell m}] = V[\tilde{a}_{\ell m}^2] = 2\langle \tilde{a}_{\ell m}^2 \rangle^2$, since the $b_{\ell m}$ are mere constants. Hence the $1\sigma$ error bar is

$$\Delta \tilde{C}_\ell \approx \frac{1}{2\ell+1} \left( \sum_{m=-\ell}^{\ell} V[\tilde{C}_{\ell m}] \right)^{1/2} \approx \left( \frac{2}{2\ell+1} \right)^{1/2} (C_\ell + b_\ell). \tag{31}$$

Thus as $\ell$ increases, the error bars will typically first decrease due to the growing number of independent $m$-modes, and then gradually start increasing again around the scale corresponding to the position errors as $N_\ell^{\rm eff}$ eventually approaches zero, making $b_\ell$ explode.

### 2.3.1 Beam function

We model the BATSE beam function as Fisher function (Fisher $et$ $al.$ 1987):

$$B^k(\hat{\bf r} \cdot \hat{\bf r}') = \frac{\exp\left[\sigma_k^{-2} \hat{\bf r} \cdot \hat{\bf r}'\right]}{4\pi \sigma_k^2 \sinh[\sigma_k^{-2}]}, \tag{32}$$

---

[5]The Gaussian assumption that we used for computing error bars was strictly valid only when $N_i \gg 1$ for each sub-population. However, since the BATSE 3B distribution of position errors form a smooth continuum, we expect the error bars derived from the Gaussian approximation to remain accurate anyway, as long as $N_\ell^{\rm eff} \gg 1$, and this is indeed numerically confirmed by Monte-Carlo simulations. We generated 1000 mock BATSE 3B catalogs with no clustering and analyzed them with the same software as the real data, extracting the multipoles $\ell \leq 40$. To within the Monte-Carlo errors (a relative error of order $1000^{-1/2} \approx 3\%$), the actual error bars were identical to those expected analytically when making the Gaussian approximation.

characterized by a location error $\sigma_k$. This is often considered by mathematicians to be the spherical version of the Gaussian distribution, and reduces to

$$B^k(\cos\theta) \approx \frac{\exp\left[-\frac{1}{2}\frac{\theta^2}{\sigma_k^2}\right]}{2\pi\sigma_k^2} \qquad (33)$$

when $\sigma_k \ll 1$ radian $\approx 60°$. The Fisher function has the advantage that it is correctly normalized (its integral over the sphere is unity) for arbitrarily large angles $\sigma_k$, which is not the case for the plane Gaussian of equation (33). It should be emphasized that although the BATSE location error distribution has usually been modeled as a Gaussian distribution, it is currently not well-enough known that one particular distribution is preferred over another, so the choice is merely one of convenience.

In the limit $\sigma_k \ll 1$ (valid for all bursts in the sample as shown in Figure 2), we have to a good approximation that

$$B_{\ell k} \approx e^{-\frac{1}{2}\sigma_k^2 \ell(\ell+1)}. \qquad (34)$$

The position uncertainties $\Delta\theta$ quoted in the BATSE 3B catalog are defined as the radius of the one sigma circle, *i.e.*, of the circle that contains $\text{erf}[1/\sqrt{2}] \approx 68\%$ of the probability. Thus in the limit $\sigma_k \ll 1$, the conversion between $\Delta\theta$ and $\sigma$ is

$$\frac{\sigma}{\Delta\theta} = \left[-2\ln\left(1 - \text{erf}\left[\frac{1}{\sqrt{2}}\right]\right)\right]^{-1/2} \approx 0.66. \qquad (35)$$

Note that the values of $\Delta\theta$ quoted in the BATSE 3B catalog do not include the systematic error contribution of $1.6°$, which is to be added to the quoted values in quadrature. This yields the distribution shown in Figure 2.

## 3  RESULTS

We have used the improved BATSE positions of the 3B catalog (Meegan et al. 1995b, 1995c) to expand the angular distribution of GRBs in terms of spherical harmonics. The 3B catalog contains 1122 bursts with known best fit positions (shown in Figure 1a) and their statistical uncertainties. In addition to statistical shifts we must also include (in quadrature) a $1.6°$ systematic uncertainty. This value is significantly lower than the $4°$ of earlier catalogs and it allows us to extend spherical harmonic analysis to $\ell \sim 50 - 60$ before localization uncertainties wash out any possible intrinsic angular power in the GRB sky map. The distribution of the actual statistical errors is shown in Figure 2

Because the sky exposure of BATSE is not uniform (Fishman et al. 1994; Meegan et al. 1995b), artificial moments are induced (e.g., Briggs *et al.* 1995). The BATSE experiment does not exclude any area of the sky, but due to blocking by the Earth and detector gaps during passages of the SAA some positions on the sky have a reduced probability for burst detection. The associated exposure map is thus best described as a semi-transparent mask. While the exposure correction is not as severe as those encountered in galaxy surveys it should and can be included in the analysis. We shall discuss the effect of uneven sampling in the next section.

## 3.1 The exposure function

Because of problems due to the loss of the spacecraft tape recorders, the absolute efficiency has not been determined since the release of the 1B data set. However, the shape of the exposure function $\bar{n}$ is essentially independent of time, and since the shape is all that matters for the present analysis, we employ the 1B estimate (Fishman *et al.* 1994). This function $\bar{n}$ depends on declination only, and is independent of right ascension. This means that in equatorial coordinates, the multipole coefficients $\bar{n}_{\ell m}$ vanish except when $m = 0$. The dominant deviation from uniformity is a quadrupole ($\bar{n}_{20}/\bar{n}_{00} \approx 8.8\%$) depletion of bursts near the equator due to the shadowing of the sky by the earth. The second largest anisotropy is a dipole moment ($\bar{n}_{10}/\bar{n}_{00} \approx 4.5\%$) towards the earth's north pole, due to the South Atlantic Anomaly, which requires disabling triggers. Compared to the shot noise, the higher multipoles ($\ell \geq 3$) are negligible ($a_{\ell 0}/a_{00} \lesssim 1\%$), but for completeness, they have nonetheless been included in our analysis.

## 3.2 The power spectrum

The power spectrum $\tilde{C}_\ell$ extracted from the BATSE 3B data set is shown in Figure 3, and as can be seen, there is no evidence of deviations from isotropy on any angular scale. What is plotted is of course the difference between two positive quantities, the power in the data minus the bias correction, according to equation (12), which is why some (unphysical) negative estimates occur. Thus if the gamma-ray bursts are in fact completely uncorrelated, we would expect the points in Figure 3 to be scattered symmetrically around zero, with roughly equal numbers above and below the horizontal axis, and about 68% within the shaded region. Since all power is by definition positive, the presence of any type of correlation would shift the distribution upwards, leading to a positive excess.

In Figure 3, we have divided the power spectrum by $4\pi$ to make the interpretation of the numbers simpler. A monopole $C_0/4\pi = 0.0001$ would simply correspond to a fluctuation of $\sqrt{0.0001} = 1\%$ in the average burst density. Likewise, $[C_\ell/4\pi]^{1/2}$ can be interpreted as the density fluctuation on the angular scale $\theta \approx 60°/\ell$.

Let us briefly comment of this factor of $60°$ and the correspondence between $\ell$ and $\theta$. From equation (34), we see that roughly speaking, a burst only probes the multipole $\ell$ if the factor $B_{\ell k}$ is of order unity, i.e., if $\sigma_k \ell \lesssim 1$. Here $\sigma_k$ is measured in radians, so since one radian is $180°/\pi \approx 57°$, this means that only bursters with a location error $\sigma \lesssim 60°/\ell$ are sensitive to the multipole $\ell$.

The size of the error bars (the height of the shaded region) in Figure 3 is readily understood from equation (31). For $\ell = 0$, we have $N_\ell^{\text{eff}} = N = 1122$, so apart from the factor of $\sqrt{2}$, the shot noise gives just the familiar Poisson variance $1/N$. As $\ell$ increases, the $(2\ell+1)$-denominator reduces the error bars, since many independent modes are being averaged. However, as $\ell$ increases beyond the scale corresponding to the typical location errors, the sharp drop in $N_\ell^{\text{eff}}$ causes the error bars to increase dramatically. Thus we cannot place strong constraints for $\ell \gg 60$ simply because there are no bursts that are better localized than $1.6°$. This effect is the reason that the actual error bars become so much larger than the "ideal world" error bars (double-hatched) that would result if there were no position errors. This is also illustrated in Figure 4, where $N_\ell^{\text{eff}}$ is plotted as a function of $\ell$. For $\ell = 30$, for instance, we are effectively only making use of about 10% of all bursts, the remainder being too poorly localized to contribute much information about the power on this small a scale. Conversely, Figure 3 also shows that for $\ell \lesssim 5$, the location errors have little impact on the error bars, confirming the results of Horack *et al.* (1993) and Briggs *et al.* (1995) for dipole and quadrupole moments.

Note that $N_\ell^{\text{eff}}$ in Figure 4 is far from being Gaussian: for small $\ell$, it falls off roughly as a Gaussian with with $\ell = 10$, but for larger $\ell$, the tail falls off much more slowly, since most of the contribution is coming from the best localized bursts. It should also be noted that since the $C_\ell$-coefficients are rotationally invariant quantities, Figure 3 would look identical if galactic rather than equatorial coordinates had been used when generating it.

### 3.3 Is the exposure function correct?

If the estimate of $\bar{n}$ (Fishman *et al.* 1994) were incorrect, this could introduce artificial signals into our power spectrum. Because of the azimuthal symmetry, this would only affect those coefficients $\tilde{a}_{\ell m}$ that have $m = 0$. These are plotted in Figure 5. Thus if the bursts are uncorrelated and the $\bar{n}$-estimate is correct, the points should scatter symmetrically around zero with about 68% of them in the shaded region, which appears to be the case. Figure 5 thus provides reassuring evidence that $\bar{n}$ has been correctly modeled. To indicate the sensitivity of this analysis, the figure also shows the dipole and quadrupole that would be expected if we had failed to correct for the above-mentioned earth-shadow quadrupole and the South Atlantic Anomaly. Since the quadrupole correction was about 9%, this shows that uncertainties in the modeling of the higher multipoles of $\bar{n}$, which are typically at least an order of magnitude smaller, will not be important compared to the $(N_\ell^{\text{eff}})^{-1/2}$-errors caused by shot noise.

### 3.4 The minimum-variance-weighted burst map

Using equation (17) and the orthogonality relation (3), we can rewrite equation (28) as

$$\tilde{a}_{\ell m} = \frac{N}{N_\ell^{\text{eff}}} \int Y_{\ell m}(\hat{\mathbf{r}}) x(\hat{\mathbf{r}}) d\Omega, \tag{36}$$

where we have defined $x$, the *smoothed burst map*, as

$$x(\hat{\mathbf{r}}) \equiv \sum_{k=1}^{N} \frac{B^k(\hat{\mathbf{r}} \cdot \hat{\mathbf{r}}_k)}{\bar{n}(\hat{\mathbf{r}})}. \tag{37}$$

Thus we see that the minimum-variance method we derived above has a very simple interpretation: apart from the overall weighting factor $N/N_\ell^{\text{eff}}$, our optimal estimates of the multipoles $a_{\ell m}$ were just the spherical harmonic coefficients of a map where *each burst is smeared out by its own beam function*, and corrected for the uneven sky exposure. This map is shown in Figure 1b and Plate 1 (upper left). A comparison of this map with that using earlier BATSE data (Hartmann 1994) shows the tremendous improvements due to the reduction of systematic position uncertainties from 4° to 1.6° and the increase in sample size.

It is quite useful for visually inspecting the data set, since it in a sense displays only the information that is really present in the data and not more. It does not mislead the eye by exaggerating the accuracy to which the burst locations are known, enabling those bursts that are well-localized to visually stand out against the background.

## 3.5 Band-pass filtered maps

Although the angular power spectrum $C_\ell$ provides a useful measure of the amount of clustering on different angular scales, it should be borne in mind that it does not contain any information about the relative *phases* of the different multipoles $a_{\ell m}$. The same can be said about the correlation function, a useful statistical quantity that has been estimated elsewhere (Meegan *et al.* 1995b, 1995c; Blumenthal 1995). The loss of phase-information means that although the power spectrum may tell us that there is extra power on some scale, it does not tell us anything about where in the sky this power is coming from — we may for instance be interested in knowing if there are any signals localized near the Large Magellanic cloud or aligned with the galactic plane. Fortunately, this type of information (which can be seen as complementary to that provided by the power spectrum) is easy to extract with the formalism developed above. We define $x_\ell(\hat{\mathbf{r}})$, the *multipole map* corresponding to multipole $\ell$, as the sky map

$$x_\ell(\hat{\mathbf{r}}) \equiv \sum_{m=-\ell}^{\ell} \tilde{a}_{\ell m} Y_{\ell m}(\hat{\mathbf{r}}), \tag{38}$$

where the estimated spherical harmonic coefficients $\tilde{a}_{\ell m}$ are those defined by equation (28). Similarly, we define the *band-pass filtered map* corresponding to a multipole range $\ell_1 \leq \ell \leq \ell_2$ as the sum of the multipole maps for the different $\ell$-values in the range. Plate 1 shows the filtered maps corresponding to $\ell = 2$ (the quadrupole), $\ell = 3 - 10$ and $\ell = 11 - 30$, respectively, and the reader is encouraged to scrutinize these images in search for any features that are spatially localized or aligned with the galactic plane — both of which would provide evidence against isotropy. The quadrupole, for instance, is neither aligned with the galactic plane nor with the equator of Earth, and as is seen in Figure 3, its amplitude is of the order that is expected from mere shot noise fluctuations.

Using the orthogonality relation, we see that apart from the shot noise correction and a proportionality constant, our multipole estimate $\tilde{C}_\ell$ is just the integral of the square of the corresponding multipole map, $\int x_\ell^2 d\Omega$. It is in this sense that the filtered maps allow us to see were the power (the fluctuations) are coming from. Also, apart from normalization issues (for instance, the density modulation in $\bar{n}$ is eliminated in the filtered maps), the smoothed burst map in Figure 1b is just an average of all the multipole maps, weighted by inverse noise level. Thus we can think of the filtered maps roughly as a decomposition of the smoothed burst map into its different frequency components, into its contributions from different angular scales.

# 4  DISCUSSION

Much of the current debate on the origin of GRBs rests upon a careful analysis of their angular and brightness distribution. Without established counterparts or other burst properties that could be used to estimate distances we do not even know their distance scale, which in turn leaves burst energetics undetermined. Building models is a challenge under such conditions. One of the most important pieces of information that we can obtain is the angular distribution of GRBs. Deviations from isotropy on some angular scale for some or all bursts could provide crucial hints to the distance scale. The lack of large anisotropies makes it very hard to retain traditional models of neutron stars in the Galactic disk. But even models that invoke a very extended halo do predict small anisotropies that should emerge eventually from the data. And while cosmological models generically result in isotropic distributions, they too may have tell-tale deviations. We may consider the small deviations due to the Earth's motion with respect to the CMB, or the granularity due to local super structures in the cosmic mass distribution. In addition, the well known angular correlations of many cosmological objects or clustering that would result from burst recurrences would lead to some deviations from isotropy. The distribution of burst positions on the sky could be the primary source of information leading to an understanding of the burster distance scale, and perhaps their nature as well.

The crucial objective of our study is thus an advanced analysis of GRB positions. There are two significant steps in this field: 1) providing accurate locations for all bursts, and 2) analyzing this position information with appropriate statistical tools. The BATSE Team has made great progress in the first area, now providing location accuracies down to $\sim 2°$ for many bursts, and $\sim 5°$ for the average burst (Meegan *et al.* 1995c). The reduction of systematic uncertainties is essential for studies of small-scale anisotropies, but it also contributes to better estimates of more global patterns that may be present in the data. The smearing of burst positions, unavoidable from the instrumental point of view, must be included in the data analysis. Ad-ditional features that must be accounted for are temporal and angular gaps in the observations. Here, we do not consider possible structure in burst arrival times but study exclusively their arrival directions. The exposure function of BATSE must and has been included in this work.

The remaining question is about selecting appropriate tools. This depends somewhat on the question we wish to address. Global anisotropies present in many Galactic burst models can be studied through low-order multipole expansion, e.g., dipole-quadrupole statistics, while clustering is generally approached with angular correlation functions or nearest neighbor distributions. Because of the larger database and the superior position accuracy of the 3B data studied here we are actually able to bridge these two distinct approaches by extending dipole and quadrupole analysis of the angular distribution of GRBs to higher order multipoles. The technique is the well known Spherical Harmonic Analysis (SHA), *i.e.*, expansion of the burster distribution in terms of spherical harmonics, $Y_{\ell m}$. As discussed above, the angular scale probed by a given harmonic is approximately $60°/\ell$, so that the expansion must be carried out to $\ell$-values in excess of 30 if we wish to reach the smearing scale of the current BATSE configuration.

If some fraction of the observed GRBs are repeat events, the sky distribution should show angular concentrations on small scales (roughly given by the beam smearing of the instrument). Evidence for burst recurrence was found in the 1B data (Quashnock & Lamb 1993a), but subsequent 2B data did not confirm this result (e.g., Meegan *et al.* 1995a).

The 3B data is greatly improved over the 2B data in its ability to test the repeater hypothesis, since

- the systematic position uncertainty has been reduced from 4° to 1.6°, and
- in addition to the overall exposure time being increased by about a year, the post-2B portion of the 3B catalog has a greater fractional exposure (livetime), which is important for repeater models in which the bursting phase of sources is less than the BATSE lifetime.

Burst recurrence is expected to generate excess correlations at $\theta = 0$, which corresponds to excess power at all multipoles[6]. Our study does show some

---
[6]From the addition theorem (16), one obtains the well-known result that

$$\langle \Delta(\widehat{\mathbf{r}})\Delta(\widehat{\mathbf{r}}') \rangle = \sum_{\ell=0}^{\infty} \left(\frac{2\ell+1}{4\pi}\right) P_\ell(\widehat{\mathbf{r}} \cdot \widehat{\mathbf{r}}') C_\ell, \quad (39)$$

*i.e.*, the $C_\ell$ are are basically the spherical Fourier coefficients of the angular correlation modes with deviations around the $2\sigma$ level, but this is by no means a significant excess of power, because only about the expected number of points deviate by about $2\sigma$ and the points are generally scattered within $1\sigma$ of no power. The data are consistent with the hypothesis of no recurrences. Particular models can be tested with SHA, and quantitative constraints on their parameter space should be derived in future studies. Here we simply conclude that burst recurrence can apparently not occur for a significant fraction of all bursts. While some actual repeaters can not be ruled out by our method, SHA suggests that their frequency must be very limited.

It is conceivable that bursts repeat once or more often on a time scale of $\sim$ months, and afterwards become dormant for a much longer period. In that situation, accumulation of bursts into a growing sample would dilute the repeater signal. When the 3B set is divided into four sets of roughly equal number of bursts (not equal time), the correlation function shows some small-angle excess at the $\sim$ 1 to $2\sigma$ level in all subsets (Meegan *et al.* 1995b; Blumenthal 1995). Adding these correlation functions together generates a noticeable, but still not highly significant, excess of burst pairs near $\sim 5°$. Our corresponding SHA analysis for the four subsets (Figure 6) also reveals this effect, but it is evident from this figure that the significance of this increase is marginal at best. In other words, SHA yields results that are consistent with those obtained by correlation function analysis. This emphasizes the fact that the SHA method now bridges the range of power estimators previously employed in GRB studies.

Angular power spectra also constrain burst models that trace the large scale structure of the universe. If GRBs trace the galaxy distribution (as neutron star binary mergers would) we expect to find angular correlations similar to those observed for galaxies or clusters of galaxies (Hartmann & Blumenthal 1989; Lamb & Quashnock 1993). However, if BATSE samples to a redshift of order unity (assuming standard cosmology and standard candles for bursts), the sparse sampling of the galaxy density (specific rate of bursts inside or near galaxies is $\sim 10^{-6}$ yr$^{-1}$) and the imperfect angular resolution reduces the expected strength of the clustering signal. With increasing sample size it will be possible to apply brightness selections, while retaining good angular resolution. So far, only the dipole- and quadrupole term have been investigated as a function of apparent source brightness, and

---

function. Correlations only at zero angular separation (before position errors are added in) corresponds to the correlation function being a Dirac delta function. Just as the regular Fourier transform of $\delta(x)$ is a constant, the power spectrum corresponding to repeaters is $C_\ell$ constant, independent of $\ell$.

interpreted in the context of Galactic GRB models (Quashnock & Lamb 1993b; Gurevich, Zharkov, Zybin, & Ptitsyn 1993, 1994) and cosmological GRB models (Hartmann, Briggs, & Mannheim 1995).

We conclude that multipole expansion of the projected distribution of GRBs does not show evidence for clustering on any angular scale. This argues against the recurrence of a large fraction of burst sources and against any source population with intrinsically strong anisotropies resulting from an intrinsically special position of the observer. The remarkable degree of isotropy of GRBs severely constrains any burst model that invokes traditional geometric features of the Milky Way (disk, bulge, or halo). If one wishes to retain the Galactic origin hypothesis by introducing very extended halo distributions, it seems that these populations can not contribute significantly to the dynamics of the Galaxy (those that do are all known to be highly anisotropic), but must constitute a trace population. Whether high velocity neutron stars injected into the Galactic halo can indeed provide the necessary isotropy remains to be determined, and model builders should verify that the proposed spatial distributions indeed generate essentially zero angular power on all scales, as our analysis suggests. The currently fashionable paradigm of cosmological bursts now passes this test, but eventually some deviations from isotropy are expected, and spherical harmonic analysis is a tool well suited to detect such deviations.

The authors would like to thank Ted Bunn for help with Hammer-Aitoff projections, and G. R. Blumenthal for valuable suggestions on the effects of beam smearing. This work has been partially supported by European Union contract CHRX-CT93-0120, Deutsche Forschungsgemeinschaft grant SFB-375 and NASA grant NAGW 5-1578. D.H.H. is grateful to J. Trümper for hospitality and support of a summer visit to the MPE, where this study was initiated, and to the staff and scientific members of the ITP at UCSB for their hospitality and support during the final stages of this project. Work carried out at the ITP was supported in part through NSF grant PHY94-07194.

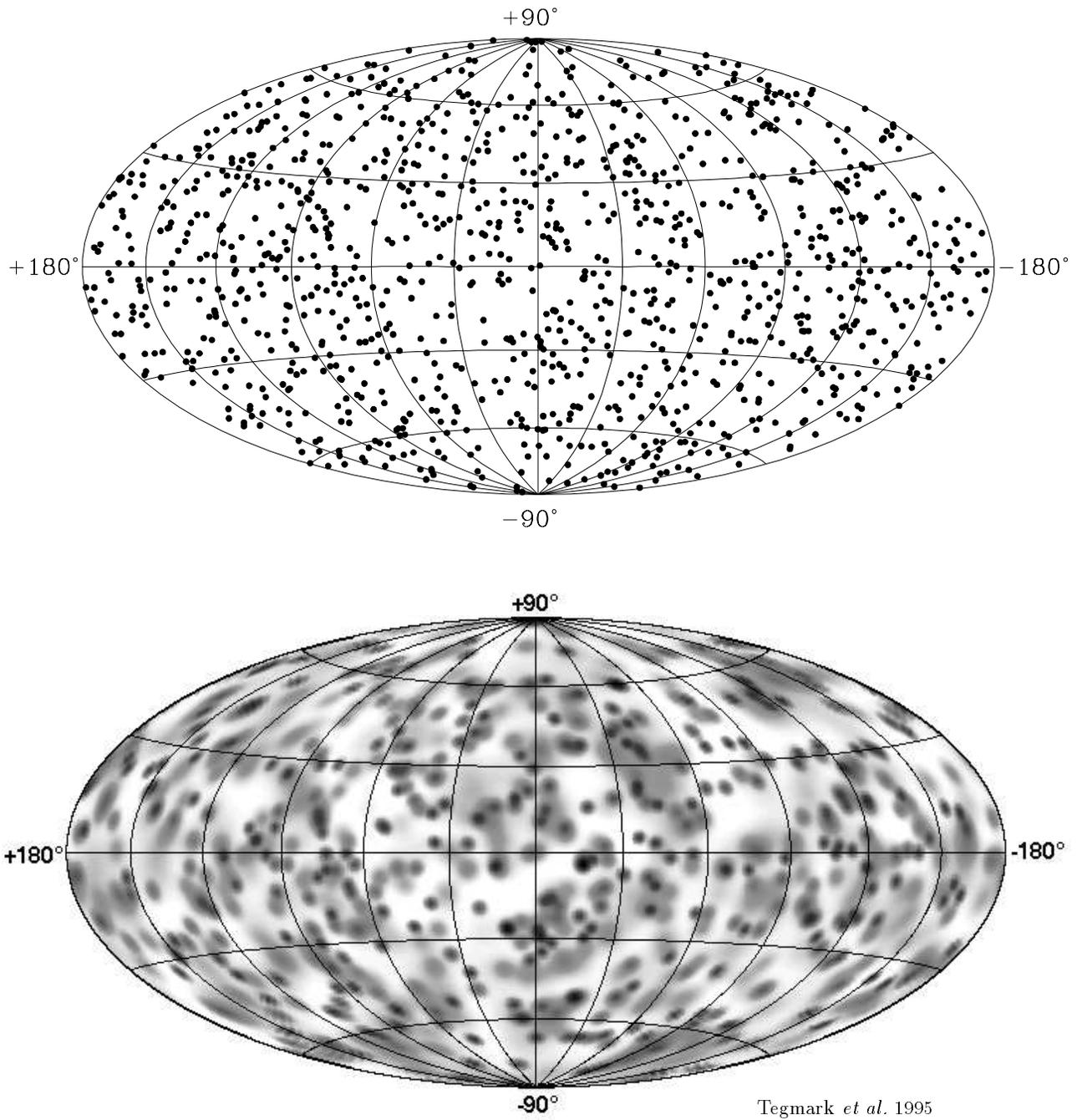

Figure 1: The BATSE 3B data set and the smoothed burst map.
The measured locations of the BATSE 3B sample of 1122 gamma-ray bursts
are shown in Hammer-Aitoff projection in galactic coordinates (Figure 1a,
top), and with each burst smeared out by an amount corresponding to the
uncertainty in its position (Figure 1b, bottom).

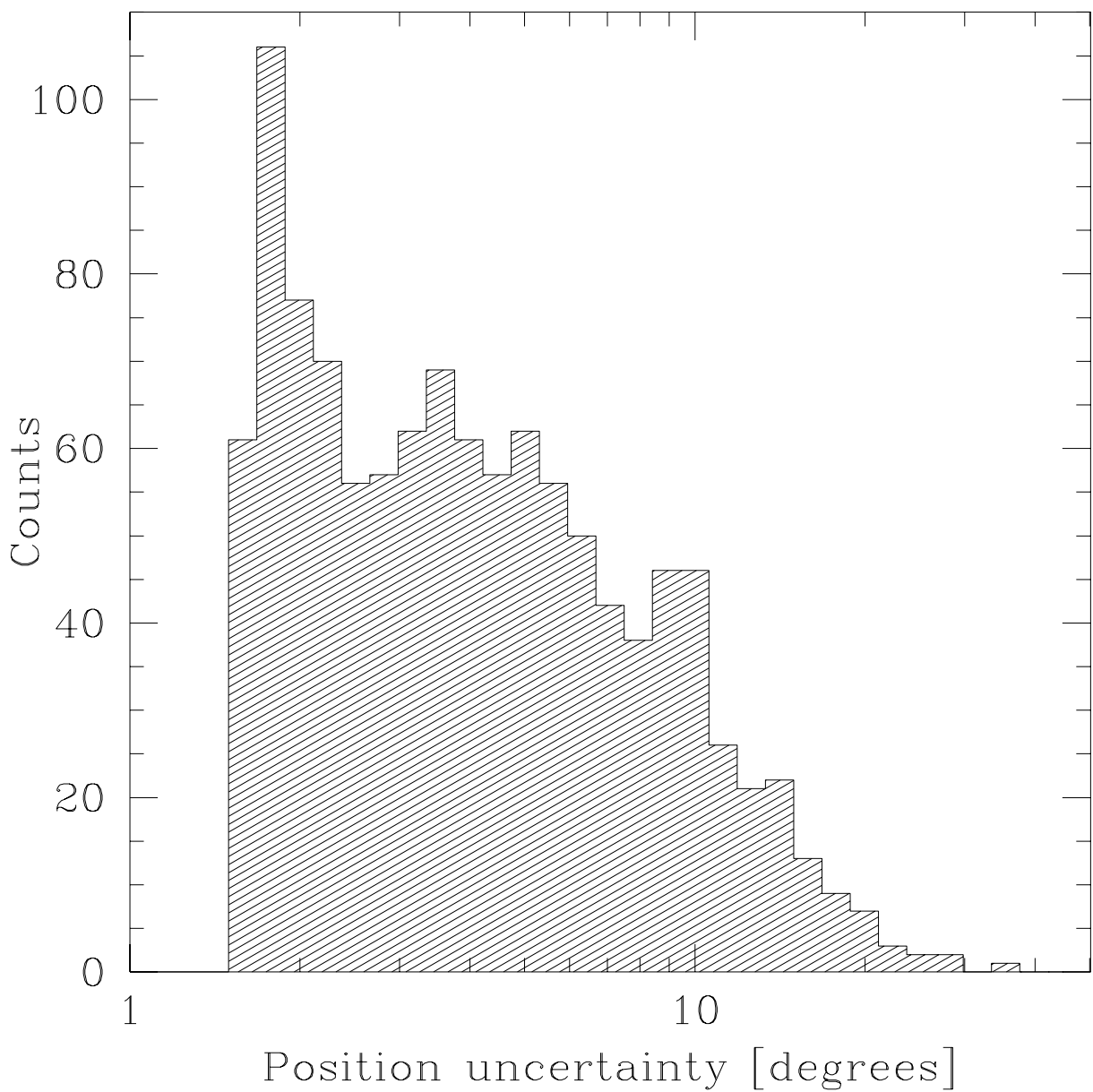

Figure 2: Position errors.
A histogram of the position errors $\Delta\theta$ is shown for the BATSE 3B sample of 1122 gamma-ray bursts. The 1.6° systematic errors are included here, added to the statistical errors in quadrature.

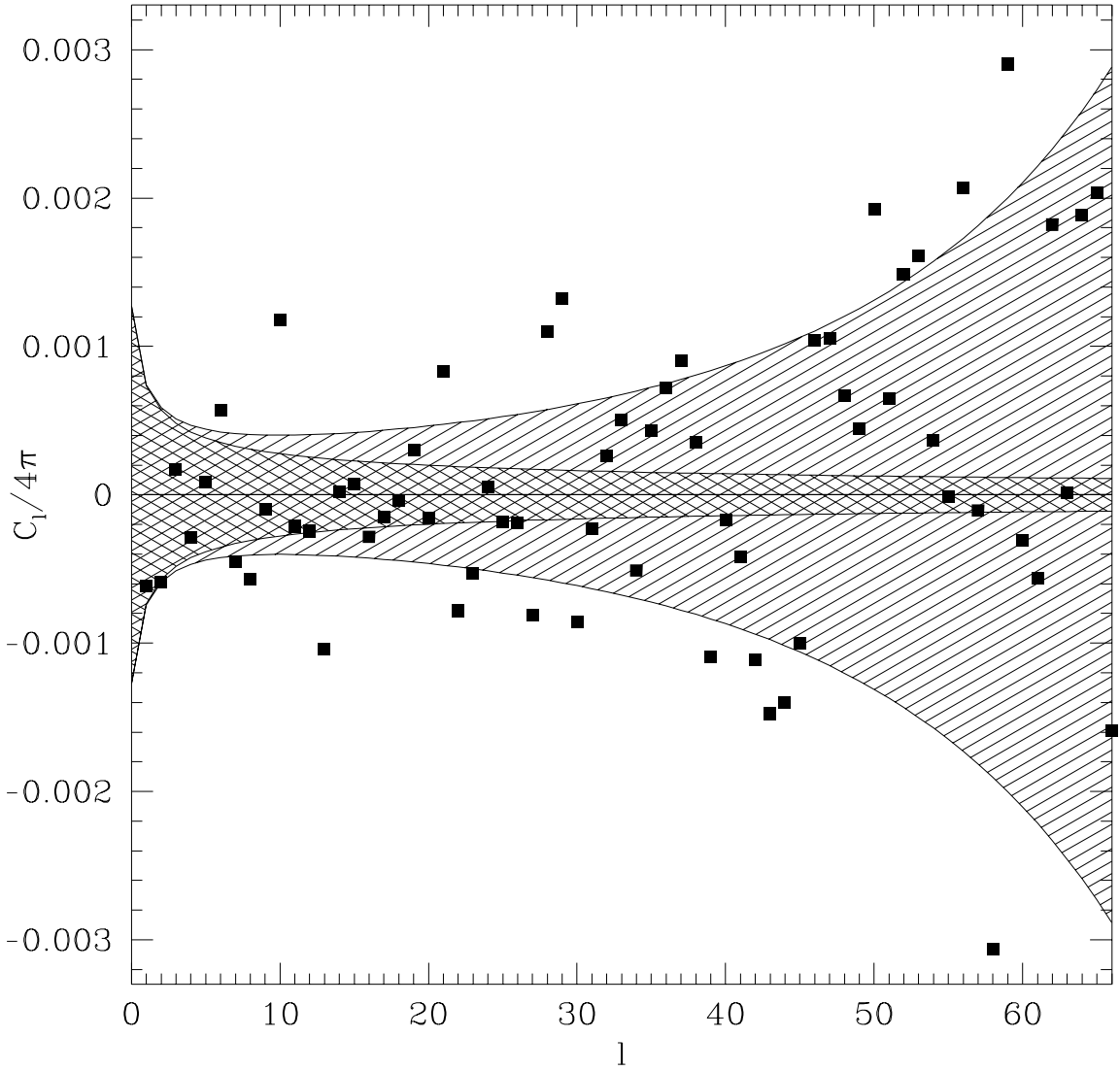

Figure 3: The shot-noise corrected angular power spectrum.
The solid squares show the multipoles estimated from the BATSE 3B data set with minimum-variance burst weighting and shot noise removed (this is why unphysical negative values occur). The shaded region shows the $1\sigma$ shot noise error bars, so if there is no clustering whatsoever, about 68% of the squares would be expected to fall within this region, symmetrically distributed abound zero. Any type of clustering would drive the points upward, leading to more points above zero than below. The double-shaded region shows what the error bars would be if there were no position errors.

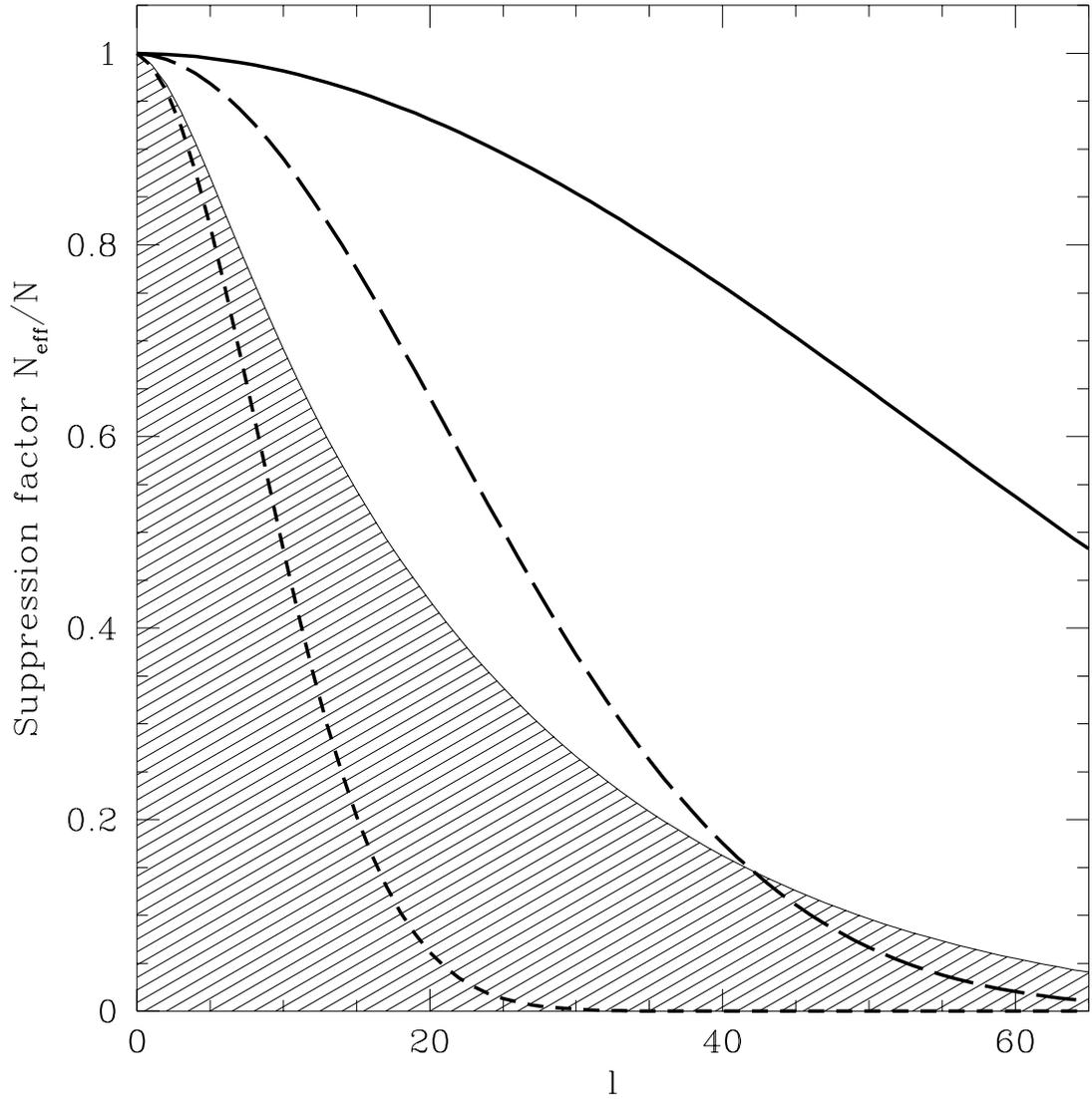

Figure 4: The effect of position errors.
The factor by which fluctuations are suppressed by the effect of position errors, $N_\ell^{\rm eff}/N$, is plotted as a function of multipole $\ell$. Our method corrects for the smearing by dividing by this suppression factor, which is the reason that the error bars in Figure 3 explode for large $\ell$. The suppression factor for the real data (shaded) is compared with the hypothetical situation where all bursts have the same position errors $\Delta\theta$, taken to be 1.6° (solid line), 4° (long-dashed line) and 10° (short-dashed line).

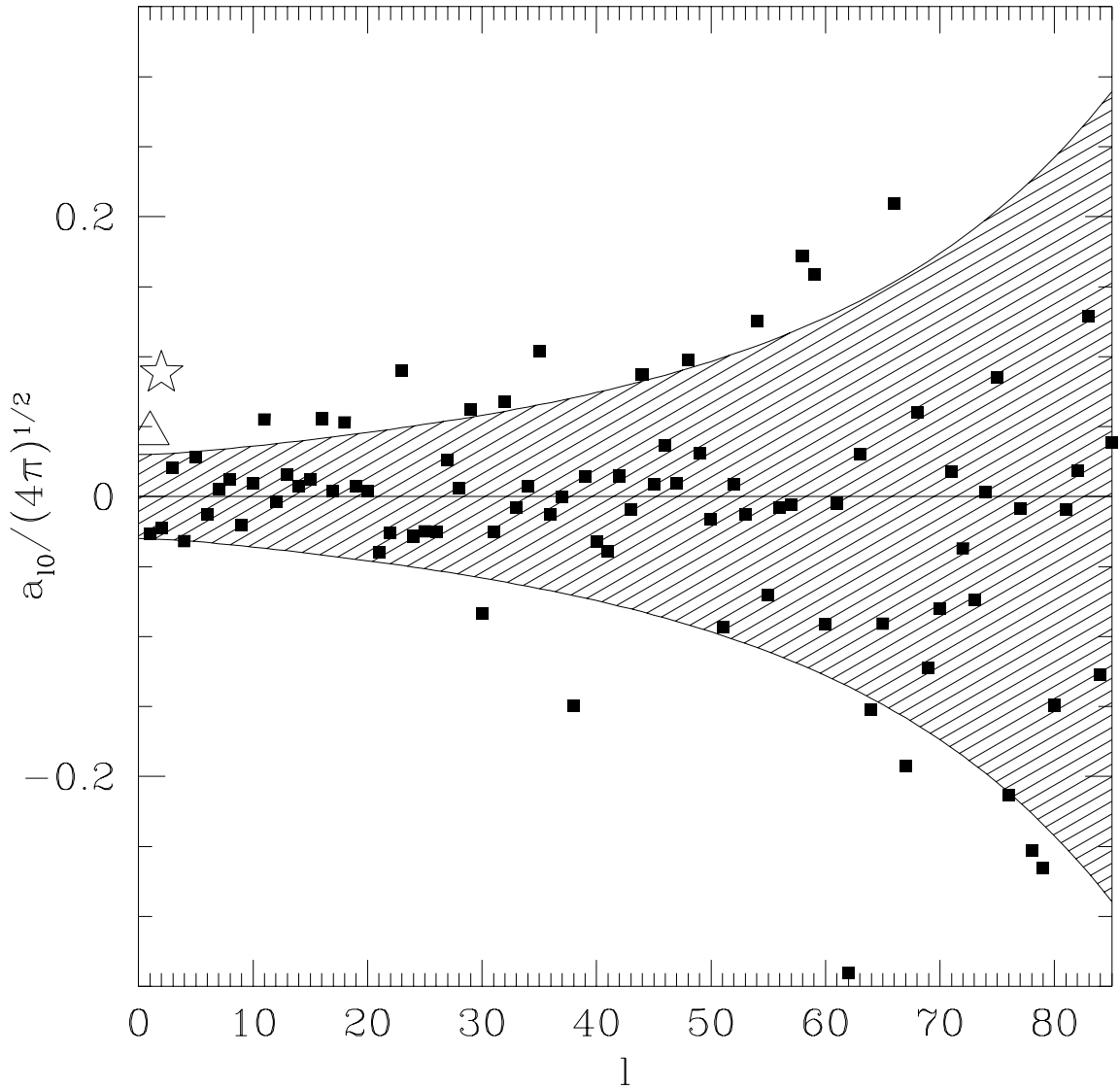

Figure 5: The multipole coefficients with $m = 0$.
The multipole coefficients $a_{l0}$ in equatorial coordinates, corresponding to fluctuations independent of right ascension, are shown (solid squares) together with the $1\sigma$ region expected from shot noise alone. For any isotropic fluctuations, the distribution should be symmetric around zero. The triangle and the star show the effect that the South Atlantic Anomaly and earth-shadowing would have if they were not taken into account in $\bar{n}$.

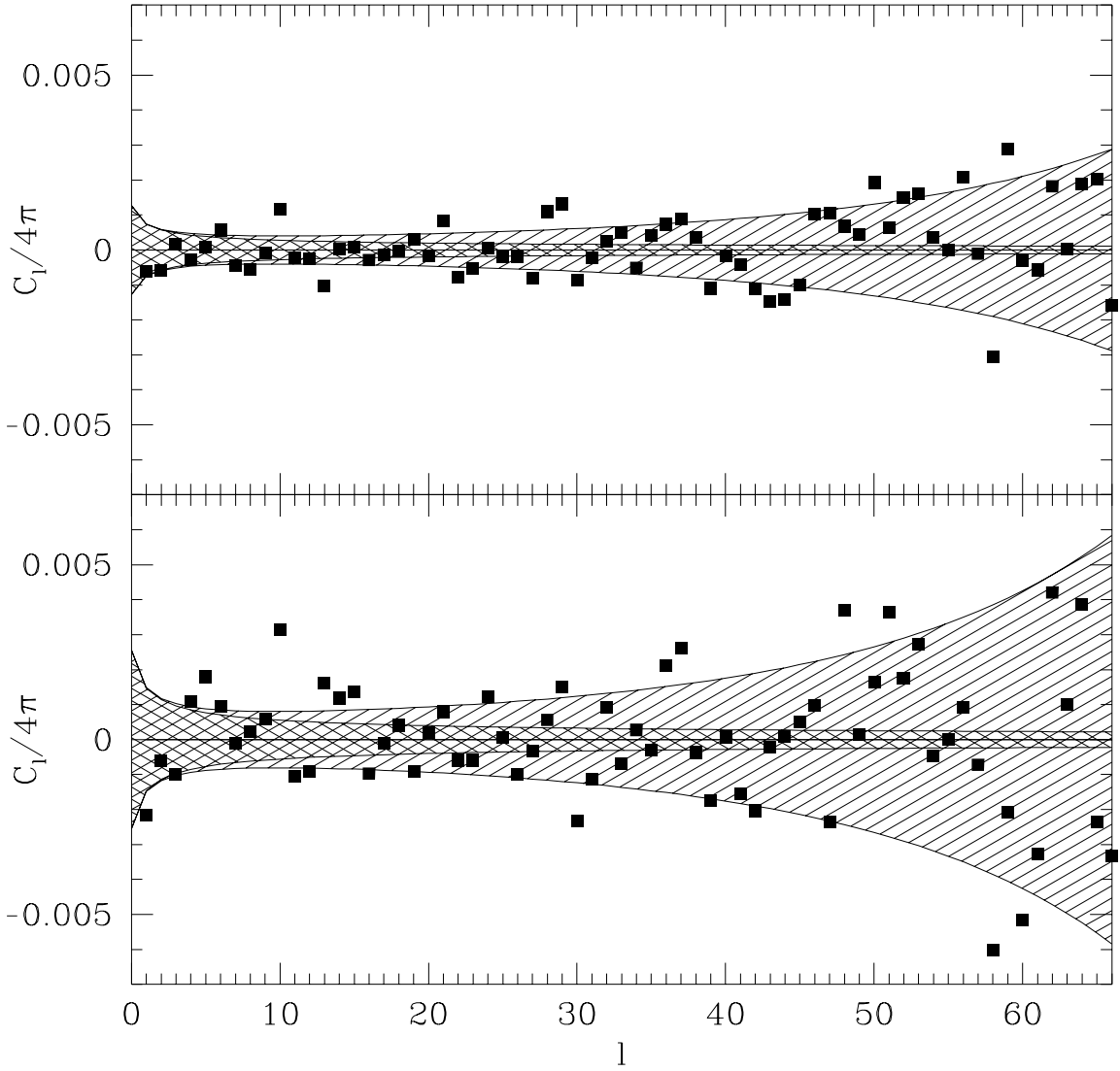

Figure 6: Angular power spectrum when grouping into quarters
The solid squares in the bottom plot show the multipoles estimated as in Figure 3, except that the data has been split into four sequential quarters according to when the burst occurred. Each square shows the average of the four estimates of that multipole, and the corresponding error bars are seen to be twice as large as in Figure 3, shown above for comparison. The slight apparent excess of power in the bottom figure is consistent with the correlation function analysis of Meegan *et al.* (1995b, 1995c), which finds weak clustering when the data is time-binned into four quarters.

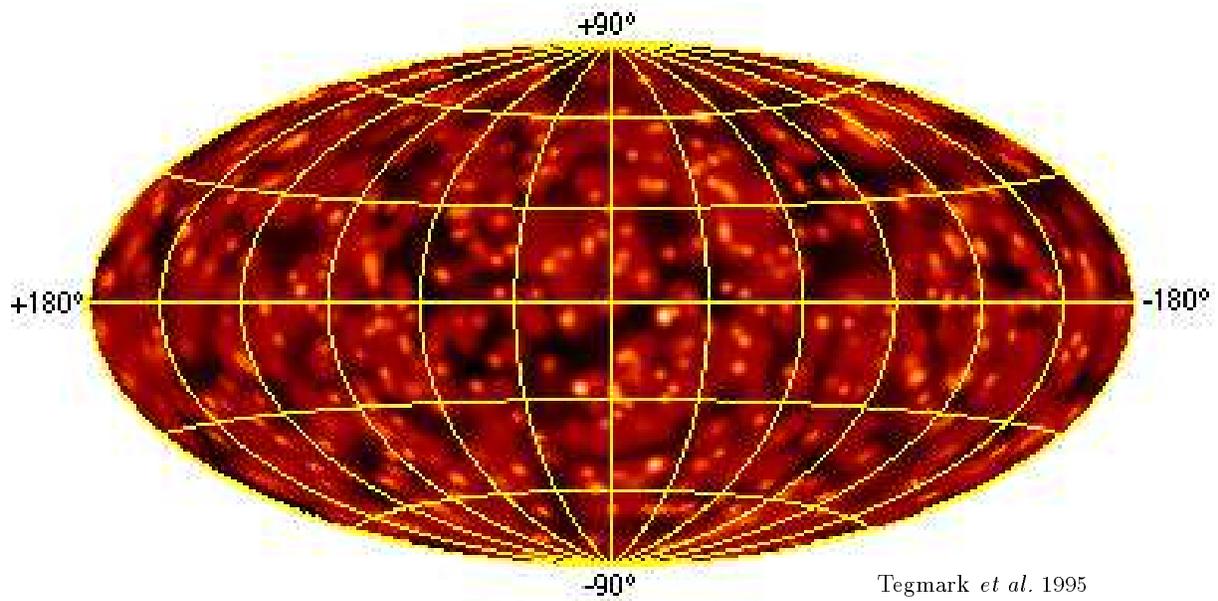

Tegmark *et al.* 1995

Figure 7: The smoothed burst map.
[This figure is a low-resolution version of the upper left quarter of Plate 1.]
An all-sky map of the 1122 gamma-ray bursts in the BATSE 3B data set
is shown, in Hammer-Aitoff projection, with each burst smeared out by an
amount corresponding to the uncertainty in its position.

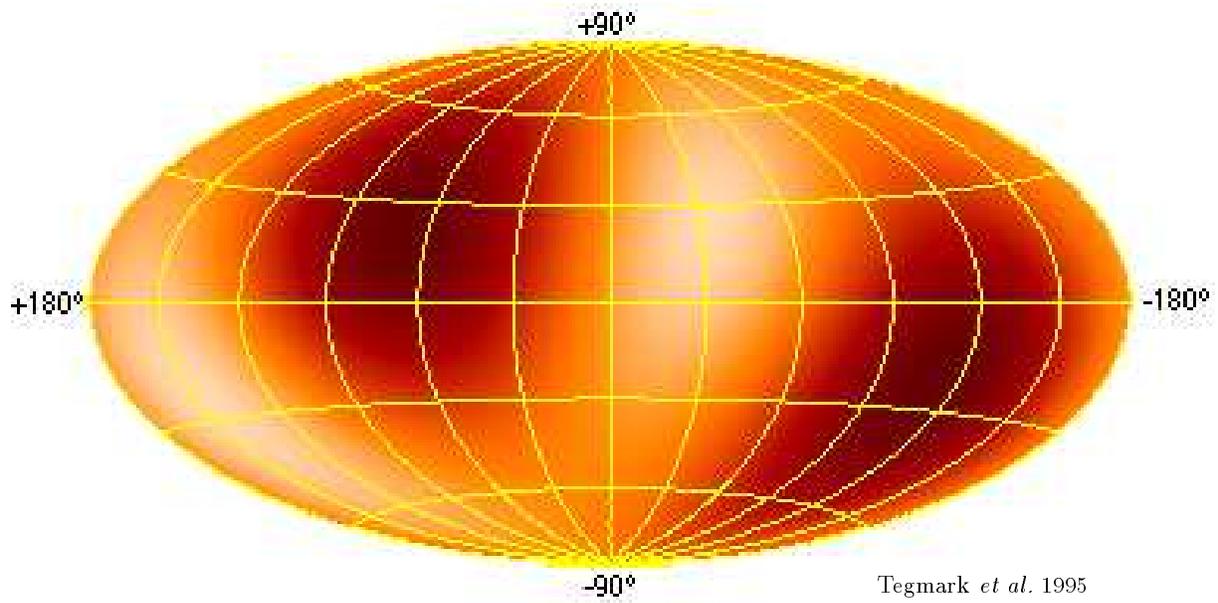

Figure 8: The BATSE 3B quadrupole.
[This figure is a low-resolution version of the upper right quarter of Plate 1.]
The multipole map for $\ell = 2$ (the quadrupole) is shown in Hammer-Aitoff projection, in galactic coordinates.

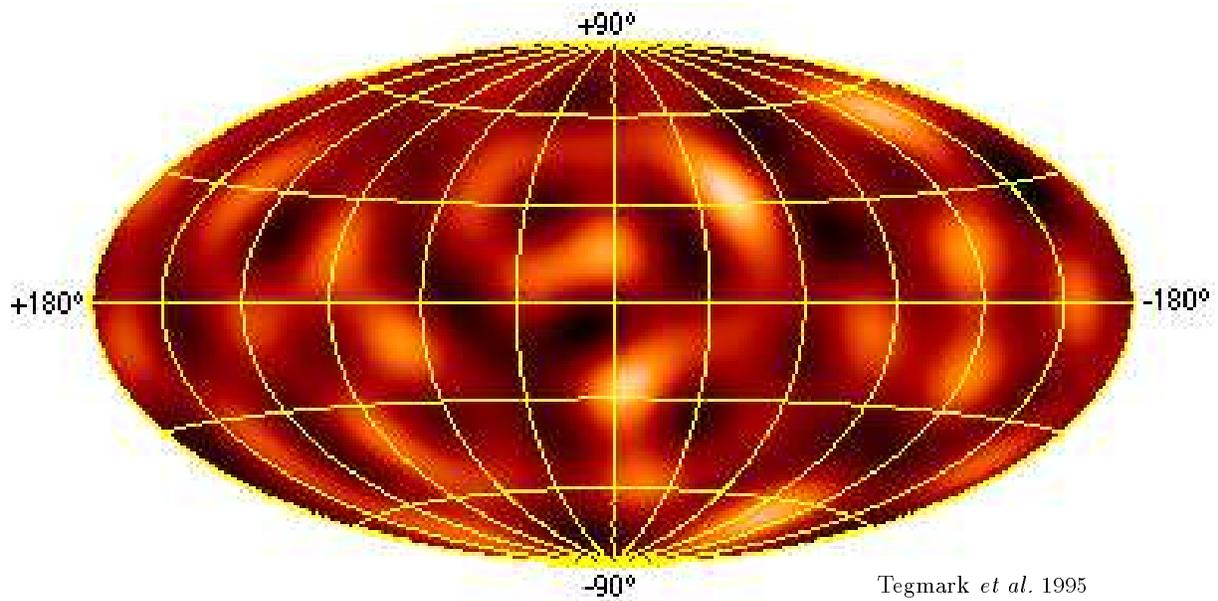

Figure 9: Fluctuations on intermediate scales.
[This figure is a low-resolution version of the lower left quarter of Plate 1.]
The band-pass filtered map for the multipole range $3 \leq \ell \leq 10$ is shown in
Hammer-Aitoff projection, in galactic coordinates.

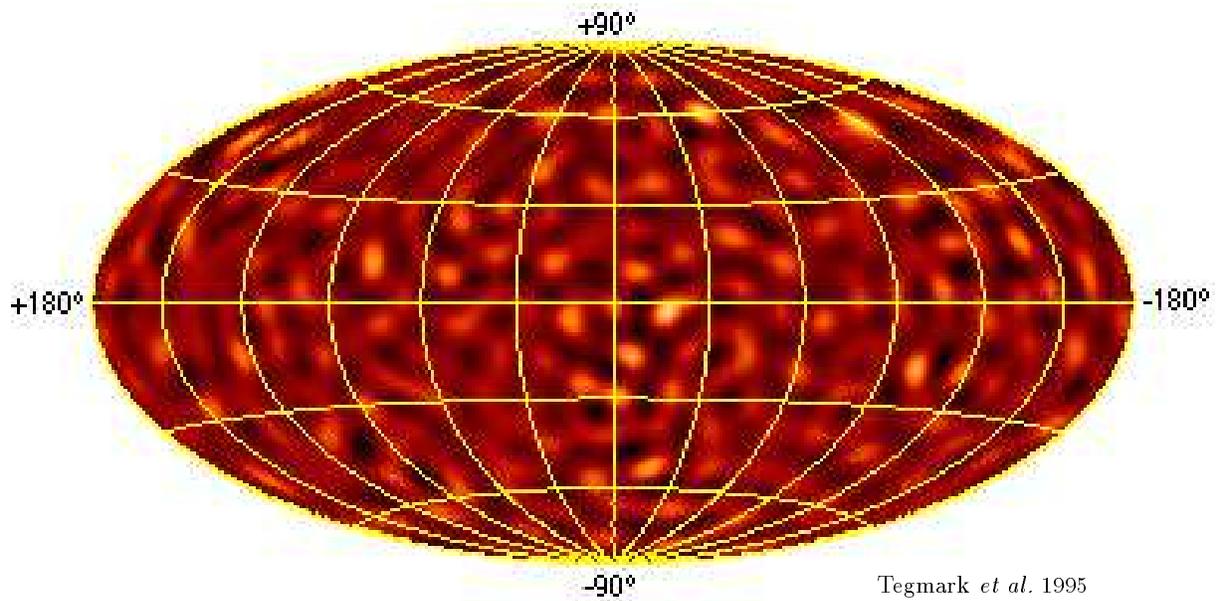

Figure 10: Fluctuations on small scales.
[This figure is a low-resolution version of the lower right quarter of Plate 1.]
The band-pass filtered map for the multipole range $11 \leq \ell \leq 30$ is shown in Hammer-Aitoff projection, in galactic coordinates.